\documentclass[12pt]{article}

\newif\iffigs\figstrue

\usepackage{epsfig,latexsym}
\usepackage{amsmath}
\usepackage{color}
\usepackage{verbatim}
\usepackage{mathrsfs}
\usepackage{amssymb}
\usepackage{slashed}
\usepackage{float}
\usepackage{booktabs}% for top-,mid- & bottomrule
\usepackage{caption}
\usepackage[mathscr]{euscript}
\usepackage{color}
\usepackage{graphicx}
\usepackage{epsf}
\usepackage{graphicx,epsfig}
\usepackage{amsmath}
\usepackage{multirow}
\usepackage{amsmath, amsthm, amssymb}
\usepackage{epsfig}
\usepackage{cite}
\usepackage{color,colordvi}

\DeclareMathAlphabet{\mathpzc}{OT1}{pzc}{m}{it}

 \csname
@addtoreset\endcsname{equation}{section}

\def\gz0{\gamma^{0}}

\def\beq{\begin{equation}}
\newcommand{\eeq}[1]{\label{#1}\end{equation}}
\def\bea{\begin{eqnarray}}
\newcommand{\eea}[1]{\label{#1}\end{eqnarray}}
\def\ba{\begin{array}}
\def\ea{\end{array}}
\def\bec{\begin{center}}
\def\ec{\end{center}}
\def\ba{\begin{align}}
\def\ena{\end{align}}

\def\12{\frac{1}{2}}

\newcommand{\I}{\mathrm{Im}}

\newcounter{hran}

\makeatletter
\renewcommand\section{\@startsection {section}{1}{\z@}%
                               {-3.5ex \@plus -1ex \@minus -.2ex}%
                               {2.3ex \@plus.2ex}%
                               {\normalfont\large\bfseries}}
\makeatother

\vspace{0.5cm}

\newcommand{\bi}{\begin{itemize}}
\newcommand{\ei}{\end{itemize}}

\begin{document}

\begin{flushright}
CERN-PH-TH/2015-050\\
%{\today}
\end{flushright}

\vspace{15pt}

\begin{center}

%%%%%%%%%%%%%%%%%%%%%%%%%%%%%%%%%%%%%%%%%%%%%%%%%%%%%%%%%%%%%%%%%%%%

{\Large\sc Doubly Self--Dual Actions \vskip 12pt in Various Dimensions}\\

%%%%%%%%%%%%%%%%%%%%%%%%%%%%%%%%%%%%%%%%%%%%%%%%%%%%%%%%%%%%%%%%%%%%

\vspace{35pt}
{\sc S.~Ferrara${}^{\; a,b,c}$, A.~Sagnotti${}^{\; a,d}$ and A.~Yeranyan${}^{\; e,b}$}\\[15pt]

{${}^a$\sl\small Department of Physics, CERN Theory Division\\
CH - 1211 Geneva 23, SWITZERLAND \\ }
e-mail: {\small \it sergio.ferrara@cern.ch}
\vspace{6pt}

{${}^b$\sl\small INFN - Laboratori Nazionali di Frascati \\
Via Enrico Fermi 40, I-00044 Frascati, ITALY}\vspace{6pt}

{${}^c$\sl\small Department of Physics and Astronomy \\ U.C.L.A., Los Angeles CA 90095-1547, USA}\vspace{6pt}

{${}^d$\sl\small
Scuola Normale Superiore and INFN\\
Piazza dei Cavalieri \ 7\\I-56126 Pisa \ ITALY \\}
e-mail: {\small \it sagnotti@sns.it}\vspace{6pt}

{${}^e$\sl\small Centro Studi e Ricerche Enrico Fermi\\
Via Panisperna 89A, 00184, Roma, Italy\\}
e-mail: {\small \it ayeran@lnf.infn.it}

%%%%%%%%%%%%%%%%%%%%%%%%%%%%%%%%%%%%%%%%%%%%%%%
\vspace{8pt}

%%%%%%%%%%%%%%%%%%%%%%%%%%%%%%%%%%%%%%%%%%%%%%%
\vspace{25pt} {\sc\large Abstract}
\end{center}
The self--duality of the $N=1$ supersymmetric Born--Infeld action implies a double self--duality of the tensor multiplet square--root action when the scalar and the antisymmetric tensor are interchanged via Poincar\'e duality. We show how this phenomenon extends to $D$ space--time dimensions for non--linear actions involving pairs of forms of rank $p$ and $D-p-2$. As a byproduct, we construct a new two--field generalization of the Born--Infeld action whose equations of motion are invariant under a $U(1)$ duality. In these systems, the introduction of Green--Schwarz terms results in explicit non--linear mass--like terms for dual massive pairs.

%%%%%%%%%%%%%%%%%%%%%%%%%%%%%%%%%%%%%%%%%%%%%%%
\vfill
%\line(1,0){250}\\

\noindent

\setcounter{page}{1}
\baselineskip=20pt

\pagebreak

\newpage
\section{Introduction}\label{sec:intro}
The partial breaking of extended supersymmetry is closely linked to the physics of branes, as was originally observed in \cite{HLP}.
Different realizations of $N=2$ Supersymmetry spontaneously broken to $N=1$ \cite{BG,BG2,RT} give rise to physically different non--linear Lagrangians whose propagating degrees of freedom are an $N=1$ vector multiplet, or alternatively an $N=1$ tensor multiplet or chiral multiplet, where the latter two options are dual to one another. For instance, the Supersymmetric Born--Infeld action \cite{deser,CF} inherits from its bosonic counterpart, which is the standard Born--Infeld action \cite{BI}, its self--duality. However, the tensor and chiral multiplet actions enjoy a different type of duality, where an antisymmetric tensor is turned into a scalar and/or vice versa. This new duality, which we call double self--duality, leads to three dual Lagrangians, depending on whether the two spinless massless degrees of freedom are described via a scalar and antisymmetric tensor, two scalars or two antisymmetric tensors. While in the first case the action turns out to be doubly self--dual \cite{BG2}, in the other cases a double duality maps one action into the other. The three actions are also connected by a single duality affecting only one of the two fields.

As pointed out in \cite{BG,RT}, in four dimensions the close connection between the Supersymmetric Born--Infeld action and the non--linear tensor multiplet action stems from similarities between the superspace $N=1$ constraints underlying the two models. Indeed, introducing an $N=1$ vector multiplet chiral field strength $W_\alpha=\overline{D}^{\,2} \, D_\alpha V$ ($\overline{D}_{\dot{\alpha}}W_\alpha=0$) and the corresponding object $\psi_\alpha = D_\alpha L$ ($D_\alpha \, \psi_\beta=0$) for a linear multiplet ($D^{\,2}L = {\overline{D}}^{\,2}L=0$), the non--linear actions in the two cases are determined by the non--linear constraints
\beq
X \, = \, - \ \frac{W_\alpha^{\,2}}{\mu \ - \ {\overline{D}}^{\,2} \,\overline{X}}  \qquad {\rm and } \qquad X \, = \, - \ \frac{\psi_\alpha^{\,2}}{\mu \ - \ {D}^{\,2} \,\overline{X}}\ ,
\eeq{1}
where $\mu$ is a parameter with mass--square dimension that sets the supersymmetry breaking scale. $X$ is chiral in the first case, since $W_\alpha$ is a chiral superfield, and is antichiral in the second, since $\psi_\alpha$ is an antichiral one. However, the highest components of the equations are identical, provided one maps the complex field
\beq
G_+^{\,2} \ = \ F^{\,2} \ + \ i\, F\, \widetilde{F}
\eeq{2}
of the Born--Infeld action into the complex field
\beq
\left( \partial \phi \right)^2 \ - \ H_{\mu}^{\,2}  \ + \ i\, H^{\mu}\, \partial_\mu \phi
\eeq{3}
of the linear multiplet action, where
\beq
H_\mu \ = \ \frac{1}{3!} \ \epsilon_{\mu\nu\rho\sigma}\, \partial^\nu\, B^{\rho\sigma} \ .
\eeq{4}
For both systems, the non--linear Lagrangian is proportional to the $F$--component of the chiral(antichiral) superfield $X$. This is subject to the constraint in \eqref{1}, which implies in both cases its nilpotency, $X^2=0$ \cite{nilpotent}.

Supergravity models of inflation based on nilpotent superfields, starting from the Starobinsky model constructed in \cite{ADFS}, were recently proposed \cite{FKL}, and were found to place interesting restrictions on model building \cite{KL,DZ}. Nilpotent superfields are also closely related to ``brane supersymmetry breaking'' in String Theory \cite{bsb,dm}, to the KKLT \cite{KKLT} construction \cite{KW} and to supersymmetry breaking in de Sitter vacua \cite{KvP}.

In this note we generalize the setup to pairs of forms in $D$ dimensions
having complementary field strengths
\beq
H_{p+1} \ = \ d\, B_p \, \qquad V_{D-p-1} \ = \ d\, A_{D-p-2}  \ ,
\eeq{5}
so that one can write the geometrical term
\beq
H_{p+1} \wedge V_{d-p-1} \ .
\eeq{6}
General duality properties for massless higher--form gauge fields, including some of the models considered here, were previously studied by Kuzenko and Theisen in \cite{kst}.

The tensor multiplet non--linear Lagrangian enjoys a \emph{double self--duality}, in the following sense. To begin with, one can either dualize the scalar into a two--form or the two--form into a scalar. In both cases, the resulting Lagrangian involves two fields of the same type and is symmetric under their interchange. Hence, double self--duality is guaranteed by the symmetry, since two successive Legendre transforms yield the identity. This result therefore applies to the class of non--linear Lagrangians related to the Born--Infeld one and brought about by Supersymmetry, but also, in principle, to more general ones. All these systems can be formulated in terms of a pair of complex Lagrange multipliers \cite{RT,ADT}: the first provides a non--linear constraint, whose solution determines the value of the second, which in its turn determines the non--linear action in a square--root form. The pattern is along the lines of what was discussed in detail in \cite{FPS,FPSSY}.

The paper is organized as follows. In Section \ref{sec:Lagrangian} we derive the doubly self--dual action that generalizes to $D$ dimensions the tensor multiplet actions and contains a pairs of field strengths of rank $(p+1)$ and $(D-p-1)$ (in particular, the four--dimensional tensor multiplet is recovered for the two choices $p=0$ or $p=2$). The square root action includes a quartic term, which is the square of the geometrical coupling in eq.~\eqref{6}. The action obtained after a single duality is then presented. It is again of square root form, but involves two field strengths of identical rank, which can be either $p+1$ or $D-p-1$, and now the quartic term takes a non--geometrical universal form,  which vanishes for two identical fields. This action exhibits a manifest $U(1)$ symmetry.

In Section \ref{sec:4Dduality} we elaborate on the four--dimensional case, where for $p=0,2$ the symmetric action becomes a Nambu--Goto--like determinant, a property that does not hold for the mixed tensor--multiplet action. In addition, we show that for $p=1$ this construction leads to a pair of two--field actions, which are displayed in eqs.~\eqref{A11} and \eqref{A13} and differ in their quartic terms. The former is a generalization of the Born--Infeld action and has an ``electric--magnetic'' $U(1)$ duality, while the latter has a manifest $U(1)$ ``electric'' symmetry. Abiding to a common practice, we call a continuous duality transformation of the field equations ``electric'' if it does not mix electric and magnetic field strengths, or ``electric--magnetic'' if it does. For $p$ odd(even) the transformations have respectively diagonal(off--diagonal) embeddings in $Sp(2n,R)$ \cite{GZ} ($SO(n,n)$ \cite{ONN}).

In Section \ref{sec:higherDdualities} we consider the analogs of these four--dimensional two--field systems that exist for $D=2(p+1)$ and the corresponding continuous dualities. We show that for $p$ odd there is always a single $U(1)$ duality, while for $p$ even this extends to a $U(1) \times U(1)$ symmetry, so that both theories of eqs.~\eqref{B1} and \eqref{B2} realize the maximal continuous duality. An interesting example concerns a pair of two--form gauge fields, which occur in the K3 reduction of the type--IIB superstring \cite{AGW} are also ubiquitous in six--dimensional orientifold vacua \cite{orientifolds,orientifolds2}, for which these non--linear actions might play a role in connection with the breaking of supersymmetry. Non--linear theories in even dimensions with maximal duality were previously considered in \cite{ABMZ}.

In Section \ref{sec:massive} we propose massive generalizations of the doubly--self--dual Lagrangians, following \cite{FS}. In this case one can start from the action for a pair of massless $p$--form gauge fields and add a Green--Schwarz term \cite{GS} involving a pair of $(D-p-1)$--form gauge fields, together with a corresponding non--linear action for a pair of $(D-p)$--form field strengths. Going to a first--order form and integrating out the $(D-p)$--form field strengths, one is led to an action containing a non--linear curvature term for the $(p+1)$--form field strengths and a non--linear mass term for the two $p$--form gauge fields inherited from the original non--linear
action of the two dual $(D-p-1)$--form gauge fields. For $D=2(p+1)$ we also present an alternative kinetic Lagrangian involving a geometric quartic term, and in a similar fashion in $D=2p$ we present a Lagrangian with a geometric quartic coupling in the mass--like terms. Finally, we discuss the alternative option of coupling together two non--linear Lagrangians for $(p+1,D-p-1)$ and $(p,D-p)$ form field strengths, and using the $p$--form gauge field to give mass to the $(p+1)$--form one. The end result is a Lagrangian involving one massive field and two massless ones, and now the mass-like term for one field combines with the kinetic--like terms of one of the massless ones. The paper ends in Section \ref{sec:conclusions} with some concluding remarks.

%%%%%%%%%%%%%%%%%%%%%%%%%%%%%%
\section{Massless dualities} \label{sec:Lagrangian}
%%%%%%%%%%%%%%%%%%%%%%%%%%%%%%
Let us begin by considering the Lagrangian
\beq
{\cal L}=\mu^2 \,\left[\, 1\ -\ \sqrt{1\, +\, \frac{X}{\mu^2}\, -\, \frac{Y^2}{\mu^4}} \ \right]\ ,
\eeq{2.1}
in $D$ dimensions and with a mostly positive signature, which was also proposed in \cite{kst}. 
In form language \footnote{The $\star$ outside the brackets converts a top form into a zero--form, a step that is clearly necessary to describe these non--linear actions.}
\bea
X&=&\ - \ \star\bigg[ \ H_{p+1} \wedge \star \, H_{p+1} \ + \ V_{D-p-1} \wedge \star \, V_{D-p-1} \ \bigg] \ , \nonumber\\
Y&=& \star\bigg[ \ H_{p+1} \wedge V_{D-p-1} \ \bigg] \ ,
\eea{2.2}
with $H_{p+1}$ a $(p+1)$--form and $V_{D-p-1}$ a $(D-p-1)$--form, and $H_{p+1}=d\,B_{p}$ and $V_{D-p-1}=d\,A_{D-p-2}$. Alternatively, in components
\bea
X&=&\frac{1}{(p+1)!}\ H^2+\frac{1}{(D-p-1)!}\ V^2\ ,\nonumber \\
Y&=&\frac{1}{(p+1)!(D-p-1)!}\ \epsilon^{a_1...a_D}\,H_{a_1...a_{p+1}}\,V_{a_{p+2}...a_{D}}\ ,
\eea{2.3}
with
\beq
H_{a_1...a_{p+1}}=(p+1)\,\partial_{\left[a_1\right.}\,B_{\left.a_2...a_{p+1}\right]} \ , \quad V_{a_1...a_{D-p-1}}=(D-p-1)\,\partial_{\left[a_1\right.}\,A_{\left.a_2...a_{D-p-1}\right]} \ .
\eeq{2.4}

One can linearize the Lagrangian (\ref{2.1}) introducing four real Lagrangian multipliers $v$, $u$, $a_1$ and $a_2$, as in \cite{RT}. The first eliminates the square root, the second reduces its content to a quadratic expression and the others linearize some ratios. All in all, one is thus led to
\beq
{\cal L}\ = \ \frac{\mu^2}{2} \ \Im\left[\ (a\,\bar{a}\,-\,2\,a) \lambda\ +\ 2\, i\, a\ -\ \frac{1}{\mu^2} \ G\,\lambda\ \right]\ ,
\eeq{2.5}
where $a=a_1-i\, a_2$, $\lambda=u+i\, v$ and $G=X-2\, i\, Y$ .

Varying in the Lagrangian \eqref{2.5} the multiplier $\lambda$ leads to
\beq
G\ +\ \mu^{\,2}\, a\, (2\, -\, {\bar a})\ = \ 0 \ ,
\eeq{2.6}
and then letting $F=\mu\, a$, $m=2\mu$ and $G_+^2=G$, one can recover the Born--Infeld equation in the form used and generalized in \cite{FPS,FPSSY},
\beq
G^2_{+}\ + \ F\, (m\,-\, {\bar F})\ =\ 0\ .
\eeq{2.7}

In order to integrate out $H_{p+1}$, let us turn to a first--order form introducing the dual $(D-p-1)$--form gauge field $U_{D-p-1}=d\,C_{D-p-2}$,
and let us add to the Lagrangian (\ref{2.5}) the term
\beq
-\, \star\bigg[ \ H_{p+1}\, \wedge\, U_{D-p-1} \ \bigg]\ .
\eeq{2.8}
Integrating out $H_{p+1}$ then turns the total Lagrangian into
\beq
{\cal L} \ = \ \frac{\mu^2}{2} \,\Im\left[(a\,\bar{a}\,-\,2\,a) \lambda+2\, i\, a\right]\ -\ \frac{(u^2+v^2)\,V_{D-p-1}^2+\,U_{D-p-1}^2\,-\,2\, u\left(V_{D-p-1}\cdot U_{D-p-1}\right)}{2\, v} \ .
\eeq{2.9}
In terms of the convenient shorthand notations
\beq
R_{r}^2= - \ \star\bigg[ \ R_{r} \wedge \star \, R_{r}  \ \bigg]
\ , \quad \left(R_{r} \cdot S_{r}\right)= - \ \star\bigg[ \ R_{r} \wedge \star \, S_{r}  \ \bigg] \ ,
\eeq{2.10}
after integrating out the auxiliary fields one finally obtains
\beq
{\cal L} \ =\ \mu^2\left[1\ -\ \sqrt{1+\frac{V_{D-p-1}^2+U_{D-p-1}^2}
{\mu^2}\,+\,\frac{V_{D-p-1}^2\,U_{D-p-1}^2\,-\,\left(V_{D-p-1}\cdot U_{D-p-1}\right)^2}
{\mu^4}}\;\right]\ .
\eeq{2.11}
Introducing the new complex $(D-p-1)$--form
\beq
W_{D-p-1}\ =\ V_{D-p-1}\ +\ i\,U_{D-p-1}\ ,
\eeq{2.12}
the Lagrangian \eqref{2.10} takes the form
\beq
{\cal L} \ =\ \mu^2\left[1\ -\ \sqrt{1\,+\,\frac{W_{D-p-1}\cdot \bar W_{D-p-1}}
{\mu^2}\,+\, \frac{\left(W_{D-p-1}\cdot\bar W_{D-p-1}\right)^2-W_{D-p-1}^2\,\bar W_{D-p-1}^2}
{4\,\mu^4}}\;\right]\ .
\eeq{2.13}
A particular case of this correspondence for $p=0,1$ and $D=3$, in which case vectors are dual to scalars, was considered in \cite{bkky}.

Let us now perform a double dualization, under which the Lagrangian (\ref{2.1}) maintains its original form. To begin with, let us add the terms
\beq
- \,\star\bigg[ \ H_{p+1}\, \wedge\, U_{D-p-1} \, +\, \ K_{p+1}\, \wedge\,V_{D-p-1} \ \bigg]
\eeq{2.14}
involving the dual gauge fields in order to move to a first--order form, and let us then integrate out $H_{p+1}$ and $V_{D-p-1}$, obtaining
\beq
{\cal L} \ =\ \frac{\mu^2}{2} \ \Im\left[(a\,\bar{a}\,-\,2\,a) \lambda\ +\ 2\, i\, a\right]\ -\ \frac{1}{2}\ \frac{v}{u^2+v^2}\ X_1\ -\ \frac{u}{u^2+v^2}\ Y_1\ ,
\eeq{2.15}
where
\bea
X_1&=&\,K_{p+1}^2 \ + \ U_{D-p-1}^2 \ , \nonumber\\
Y_1&=& \star\bigg[ \ K_{p+1} \wedge U_{D-p-1} \ \bigg]\ .
\eea{2.16}
Eliminating the axillary fields recovers the initial Lagrangian
\beq
{\cal L}\ =\ \mu^2 \,\left[\, 1\ -\ \sqrt{1\,+\,\frac{X_1}{\mu^2}\,-\, \frac{Y_1^2}{\mu^4}}\ \right]\ .
\eeq{2.17}

Let us mention here that the Lagrangian \eqref{2.14} can also be recast in a form similar to eq.~\eqref{2.5},
\beq
{\cal L} \ =\ \frac{\mu^2}{2} \ \Im\left[(b\,\bar{b}\,-\,2b) \lambda_1\ + \ 2\, i\, b\ -\ \frac{1}{\mu^2} \ G_1\,\lambda_1\right]\ ,
\eeq{2.18}
with $b=b_1-i\, b_2$, $\lambda_1=u_1+i\, v_1$ and $G_1=X_1-2\, i\, Y_1$ . Here
\beq
v_1\ =\ \frac{v}{u^2+v^2}\ ,\qquad u_1\ =\ -\ \frac{u}{u^2+v^2}\ ,
\eeq{2.19}
and moreover
\bea
b_1&=&\frac{-u^2-v^2+v+\sqrt{\left(u + a_2 v\right)^2 \left(-1 + u^2 + v^2\right) + \left(1 + (a_1 - 1)\, v\right)^2\left(u^2 + v^2\right)}}{v}\ ,\nonumber\\
b_2&=& a_2\ .
\eea{2.20}
\section{Dualities in four dimensions}\label{sec:4Dduality}

Let us now discuss the case of $D=4$ in more detail, starting from the Lagrangian containing one scalar field $\varphi$ and one tensor field $B_{\mu\nu}$. This corresponds to the choices $p=2$ or $p=0$, and
\beq
{\cal L}\ = \ \mu^2 \,\left[\,1\,-\,\sqrt{1\,+\,\frac{1}{6\,\mu^2}\ H_{\mu\,\nu\lambda}\,H^{\mu\,\nu\lambda}\,+\,\frac{1}{\mu^2}\ \partial_{\mu}\varphi\,\partial^{\mu}\varphi \, - \, \frac{1}{36\, \mu^4}\, \left(\epsilon^{\mu\,\nu\lambda\,\delta}\,H_{\mu\,\nu\lambda}\,\partial_{\delta}\varphi\right)^2}\ \right]\,,
\eeq{A1}
where $H_{\mu\,\nu\lambda}\ = \ 3\,\partial_{\left[\,\mu\right.}\,B_{\left.\nu\lambda\,\right]}$.

One can now turn $B_{\mu\nu}$ into another scalar field $\chi$ by a Legendre transformation, adding to the Lagrangian the term
\beq
-\ \frac{1}{6}\ \epsilon^{\mu\,\nu\lambda\,\delta}\,H_{\mu\,\nu\lambda}\ \partial_{\delta}\chi\ ,
\eeq{A2}
which turns it into a first--order form. It is instructive to perform these steps directly, without resorting to the introduction of auxiliary fields at intermediate stages. The field equation of $H_{\mu\,\nu\lambda}$ is then
\beq
\partial_{\delta}\,\chi \, = \,\frac{\frac{Y}{\mu^2} \ \partial_{\delta}\,\varphi \ + \ \frac{1}{6}\ \epsilon_{\mu\,\nu\lambda\,\delta}\,H^{\mu\,\nu\lambda}}{\sqrt{A}}\ ,
\eeq{A3}
where $A$ identifies the expression under the square root in eq.~\eqref{A1}, and
\beq
Y\ =\ \frac{1}{6}\ \epsilon^{\mu\,\nu\,\lambda\,\delta}\,H_{\mu\,\nu\lambda}\ \partial_{\delta}\, \varphi\ .
\eeq{A4}
Substituting into the last term of the Lagrangian gives
\beq
{\cal L}\ = \ \mu^2 \,\left[\ 1\ -\ \frac{1\ +\ \frac{1}{\mu^2}\ \partial_{\alpha}\, \varphi\ \partial^{\alpha}\, \varphi}{\sqrt{A}}\ \right]\,.
\eeq{A5}
One can now reconstruct in a similar fashion the other expressions $\partial_{\mu}\, \chi\,\partial^{\mu}\, \chi$,  $\partial_{\mu}\, \chi\,\partial^{\mu}\, \varphi$, and  finally
\beq
{\cal L}\ = \ \mu^2 \,\left[\ 1\,-\,\sqrt{1\,+\,\frac{\partial_{\mu}\,\chi\ \partial^{\mu}\,\chi+\partial_{\mu}\,\varphi\ \partial^{\mu}\,\varphi}{\mu^2}+
\frac{\left(\partial_{\mu}\,\chi\ \partial^{\mu}\,\chi\right)\,\left(\partial_{\mu}\,\varphi\ \partial^{\mu}\,\varphi\right)\,-\,
\left(\partial_{\mu}\,\chi\ \partial^{\mu}\,\varphi\right)^2}{\mu^4}}\ \right]
\eeq{A6}
or, introducing the complex field $z=\varphi+i \chi$,
\beq
{\cal L}\ = \ \mu^2 \,\left[\ 1\,-\, \sqrt{1\,+\, \frac{\partial_{\mu}\,z\ \partial^{\mu}\,\bar z}{\mu^2}\,+\,
\frac{\left(\partial_{\mu}\, z\ \partial^{\mu}\, \bar z\right)^2\,-\, \left(\partial_{\mu}\, z\ \partial^{\mu}\,z\right)\,\left(\partial_{\mu}\,\bar z\ \partial^{\mu}\,\bar z\right)
}{4\,\mu^4}}\ \right]\,.
\eeq{A7}
One can also trade $\varphi$ for an additional tensor field. To this end, one is to replace $\partial_\mu \varphi$ with a vector field $V_\mu$ in eq.~\eqref{A1} to then add to the initial Lagrangian the term
\beq
-\ \frac{1}{6}\ \epsilon^{\mu\,\nu\lambda\,\delta}\,K_{\mu\,\nu\lambda} \,V_\delta\ ,
\eeq{A8}
where $K_{\mu\,\nu\lambda}\ = \ 3\,\partial_{\left[\,\mu\right.}\,C_{\left.\nu\lambda\,\right]}$.

Varying ${\cal L}$ with respect to $V_\mu$ now yields
\beq
K_{\mu\,\nu\lambda}\ = \ \frac{\frac{Y}{\mu^2}\ H_{\mu\,\nu\lambda}\,+\,\epsilon_{\mu\,\nu\lambda\,\delta}\,V^{\delta}}{\sqrt{A}}\ ,
\eeq{A9}
and again one can reconstruct the expressions $K_{\mu\,\nu\lambda}\,K^{\mu\,\nu\lambda}$,  $K_{\mu\,\nu\lambda}\,H^{\mu\,\nu\lambda}$, obtaining eventually
\beq
{\cal L}= \mu^2 \,\left[\ 1-\sqrt{1+\frac{K_{\mu\,\nu\lambda}\,K^{\mu\,\nu\lambda}+H_{\mu\,\nu\lambda}\,H^{\mu\,\nu\lambda}}{\mu^2}+
\frac{K_{\mu\,\nu\lambda}\,K^{\mu\,\nu\lambda}\,H_{\alpha\beta\gamma}\,H^{\alpha\beta\gamma}-
\left(K_{\mu\,\nu\lambda}\,H^{\mu\,\nu\lambda}\right)^2}{\mu^4}}\ \right]\,.
\eeq{A10}

The other case of interest corresponds to $p=1$, and involves a pair of vectors $B^i_{\mu}$ and the non--linear Lagrangian
\beq
{\cal L}\, =\, \mu^2 \,\left[\ 1\, -\, \sqrt{1\,+\, \frac{1}{2\,\mu^2}\,\left(\, F^1_{\mu\,\nu}\,F^{1\;\mu\,\nu}\, +\,F^2_{\mu\,\nu}\,F^{2\;\mu\,\nu}\,\right)\,-\,\frac{1}{16\, \mu^4}\,\left(\,\epsilon^{\mu\,\nu\lambda\,\delta}\,F^1_{\mu\,\nu}\,F^2_{\lambda\,\delta}\,\right)^2}\ \right]\,,
\eeq{A11}
where $F^i_{\mu\,\nu}\ = \ 2\,\partial_{\left[\,\mu\right.}\,B^i_{\left.\nu\,\right]}$. The dualization now requires the addition to the Lagrangian of
\beq
\,-\,\frac{1}{4}\ \epsilon^{\mu\,\nu\lambda\,\delta}\,F^1_{\mu\,\nu}\,G^1_{\lambda\,\delta}\ ,
\eeq{A12}
with $G^1_{\mu\,\nu}\ = \ 2\,\partial_{\left[\,\mu\right.}\,C^1_{\left.\nu\,\right]}$, and proceeding as above one ends up with
\beq
{\cal L}\, =\,\mu^2 \,\left[\ 1\,-\,\sqrt{1\,+\,\frac{G^1_{\mu\,\nu}\,G^{1\;\mu\,\nu}\,+\,F^2_{\mu\,\nu}\,F^{2\;\mu\,\nu}}{2\,\mu^2}\,+\,
\frac{G^1_{\mu\,\nu}\,G^{1\;\mu\,\nu}\,F^2_{\alpha\,\beta}\,F^{2\;\alpha\,\beta}\,-\,
\left(G^1_{\mu\,\nu}\,F^{2\;\mu\,\nu}\right)^2}{16\mu^4}}\ \right]\ .
\eeq{A13}

Notice that the equations of motion of the Lagrangian \eqref{A11} are invariant under a $U(1)$ electric--magnetic duality, as can be seen from the fact that the constraint
\beq
\epsilon_{\mu\nu\rho\sigma} \left[- 4\ \frac{\partial {\cal L}}{\partial F^1_{\mu\nu}}\
\frac{\partial {\cal L}}{\partial F^2_{\rho\sigma}} \ + \ F^{1\,{\mu\nu}}\, F^{2\,{\rho\sigma}}
 \right] \ = \ 0 \
\eeq{A14}
holds. On the other hand, the full Lagrangian \eqref{A13} is manifestly invariant under an $U(1)$ rotation, consistently with the fact that the constraint \cite{GZ}
\beq
\epsilon_{\mu\nu\rho\sigma} \left[ \frac{\partial {\cal L}}{\partial G^1_{\mu\nu}}\ F^{2\,{\rho\sigma}} \ - \
\frac{\partial {\cal L}}{\partial F^2_{\rho\sigma}} \ G^{1\,{\mu\nu}}
 \right] \ = \ 0 \ ,
\eeq{A15}
where the preceding one is mapped by the Legendre transform, does not mix electric and magnetic components.

\section{Dualities in $D=2\,\left(p+1\right)$ }\label{sec:higherDdualities}
It is interesting to investigate continuous dualities in the general case $D=2\,\left(p+1\right)$ for the two classes of Lagrangians
\bea
{\cal L} &=&  \mu^2 \left[ \,1\ -\ \sqrt{1\,+\,\frac{(F_{p+1}^1)^2\,+\,(F_{p+1}^2)^2}{\mu^2}
\,+\,\frac{(F_{p+1}^1)^2\,(F_{p+1}^2)^2\,-\,\left(F_{p+1}^1\cdot F_{p+1}^2\right)^2}{\mu^4}}\,\right], \label{B1} \\ \nonumber \\
{\cal L}&=& \mu^2 \left[ \, 1\ -\ \sqrt{1\,+\,\frac{(F_{p+1}^1)^2+(F_{p+1}^2)^2}{\mu^2}
\,-\,\frac{\left(\star\left[F^1_{p+1} \wedge F^2_{p+1}  \right]\right)^2}{\mu^4}}\,\right] \ ,
\eea{B2}
where the precise meaning of the symbols is spelled out in eq.~\eqref{2.10}. To this end, let us also recall that in $D$ dimensions and with given ``mostly positive'' signature
\beq
\star\star\, F^i_{p+1}\ = \ (-1)^p \ F^i_{p+1} \ , \qquad F_{p+1} \wedge G_{p+1} \ = \ - (-1)^p G_{p+1} \wedge F_{p+1} \ .
\eeq{B3}

For a pair of fields, the corresponding duality groups are in general contained in the maximal compact subgroup $U(2)$ of $Sp(4,\,R)$ for $p$ odd, and in the maximal compact subgroup $U(1) \times U(1)$ of $SO(2,\,2)$ for $p$ even \cite{GZ,ONN}.
For a general theory involving $n$ $(p+1)$--form field strengths $Fî$ in $D=2(p+1)$ dimensions and their duals $G^i$, where
\beq
\widetilde{G}^i \ = \ (p+1)! \ \frac{\partial {\cal L}}{\partial F^i} \ ,
\eeq{B6}
the duality conditions read \footnote{Here, for instance, $G^i \, {\widetilde{G}}^j$, is a shorthand notation for a total index contraction. In form language $G^i \, {\widetilde{G}}^j = \star\left[ G^i \wedge G^j \right]$.}
\bea
&& G^i \, {\widetilde{G}}^j \ + \ F^i \, {\widetilde{F}}^j \ = \ 0 \ , \label{B4_1}\\
&& G^i \, {\widetilde{F}}^j \ - \ G^j \, {\widetilde{F}}^i \ = \ 0 \ ,
\eea{B4}
and will hold for a subset of the available values of $i$ and $j$. This subset must identify a subgroup of the maximal duality group, which is $U(n)$ for $p$ odd and $SO(n) \times SO(n)$ for $p$ even. With reference to what we stated in the Introduction, an electric duality transformation corresponds to diagonal matrices when embedded in $Sp(2n,R)$ or $SO(n,n)$, while an electric--magnetic transformation corresponds to off--diagonal ones. These two classes of matrices result in the two classes of constraints of eqs.~\eqref{B4} and \eqref{B4_1}.

The two maximal duality groups obtain when the conditions \eqref{B4} hold for all values of $i$ and $j$. The infinitesimal transformations for these groups are generated by the $2n \times 2n$ matrices
\bea
&& M_{\rm p \ \rm odd} \ = \left(\begin{array}{cc}
a & b \\ -\,b & a
\end{array}\right)\ , \qquad a = - a^T \ , \quad b = b^T  \ , \\
&& M_{\rm p \ \rm even} \ = \left(\begin{array}{cc}
a & b \\ b & a
\end{array}\right) \ , \qquad a = - a^T \ , \quad b = - b^T \ .
\eea{B5}
In our case $n=1,2$ and each of the theories in eqs.~\eqref{B1} and \eqref{B2} is invariant under a $U(1)$ subgroup of the maximal duality group. The constraints that hold are in the first case
\beq
 G^1 \, {\widetilde{F}}^2 \ - \ G^2 \, {\widetilde{F}}^1 \ = \ 0 \ ,
\eeq{B7}
and in the second
\beq
G^1 \, {\widetilde{G}}^2 \ + \ F^1 \, {\widetilde{F}}^2 \ = \ 0 \ .
\eeq{B8}
For the model of eq.~\eqref{B1}, the matrix $M$ takes the same form,
\beq
\left(\begin{array}{cc}
i\, \sigma_2 & 0 \\ 0 & i\, \sigma_2 \ ,
\end{array}\right)
\eeq{B9}
for both $p$ even and $p$ odd, since this $U(1)$ does not mix electric and magnetic components. On the other hand, for the model of eq.~\eqref{B2} the matrix is purely off--diagonal and takes the form
\beq
\left(\begin{array}{cc}
0 & \sigma_1 \\ - \sigma_1 & 0
\end{array}\right)
\eeq{B10}
for $p$ odd, and
\beq
\left(\begin{array}{cc}
0 & i\, \sigma_2 \\ i\, \sigma_2 & 0
\end{array}\right)
\eeq{B11}
for $p$ even.

Actually, for $p$ even there is more, since the topological term in eq.~\eqref{B1} has a $U(1)$ invariance. As a result, eqs.~\eqref{B1} and \eqref{B2} have an additional $U(1)$ symmetry, and thus satisfy corresponding constraints, which are respectively
\bea
&& G^1 \, {\widetilde{G}}^2 \ + \ F^1 \, {\widetilde{F}}^2 \ = \ 0 \ , \label{B12_1}\\
&& G^1 \, {\widetilde{F}}^2 \ - \ G^2 \, {\widetilde{F}}^1 \ = \ 0 \ .
\eea{B12}
Therefore, the maximal $U(1) \times U(1)$ duality symmetry is realized for both models. One of the two $U(1)$ factors is a manifest electric rotation, while the other is a genuine electric--magnetic duality.

Note that for $p$ odd a theory could allow, in principle, five types of duality symmetry, where the electric part is diagonal and rest is off--diagonal. This implies that the electric part is always $U(1)$ or is absent when the magnetic part is $U(1)$, so that the five cases correspond to $U(1)_{\rm electric}$, $U(1)_{\rm magnetic}$, $U(1) \times U(1)$, $SU(2)$ and $SU(2) \times U(1)$. Our Lagrangians for $p$ odd only possess a $U(1)$ duality, which is diagonal for eq.~\eqref{B1} and off--diagonal for eq.~\eqref{B2}. On the other hand, a simple example of a Born--Infeld--like Lagrangian that admits an $SU(2)$ symmetry is
\beq
{\cal L} \ = \ \mu^2 \left[ \, 1\ -\ \sqrt{1\,+\,\frac{F_{p+1}\cdot \overline{F}_{p+1}}{\mu^2}
\,-\,\frac{\left(\star \left[F_{p+1} \wedge F_{p+1} \right]\right)\,\left(\star \left[\overline{F}_{p+1} \wedge \overline{F}_{p+1} \right]\right)}{\mu^4}}\,\right] \ ,
\eeq{B13}
where $F_{p+1}=F^1_{p+1}+i F^2_{p+1}$ is a complex field strength and the symbols are defined in eq.~\eqref{2.10}. Note that the Lagrangian of eq.~\eqref{B13}, unlike that of eq.~\eqref{B2}, has a manifest $U(1)$ electric symmetry but also maintains the original electric--magnetic $U(1)$, so that in this case eqs.~\eqref{B12_1} and \eqref{B12} are simultaneously satisfied.

%%%%%%%%%%%%%%%%%%%%%%%%%%%%%%%%%%%%%%%%%%%%%%
\section{Massive dualities}\label{sec:massive}
%%%%%%%%%%%%%%%%%%%%%%%%%%%%%%%%%%%%%%%%%%%%%%

Let us start from the Lagrangian
\bea
{\cal L} &=& \mu^2\left[1\ -\ \sqrt{1\,+\,\frac{(H_{p+1}^1)^2+(H_{p+1}^2)^2}
{\mu^2}\,+\,\frac{(H_{p+1}^1)^2\,(H_{p+1}^2)^2-\left(H_{p+1}^1\cdot H_{p+1}^2\right)^2}
{\mu^4}}\;\right]  \\ \nonumber
&+& \nu^2\left[1\ -\ \sqrt{1\, +\, \frac{(F_{D-p}^1)^2+(F_{D-p}^2)^2}
{\nu^2}\, +\, \frac{(F_{D-p}^1)^2\,(F_{D-p}^2)^2\, -\, \left(F_{D-p}^1\cdot F_{D-p}^2\right)^2}
{\nu^4}}\;\right],
\eea{3.1}
where $H_{p+1}^i=d\,B_p^i$ are $(p+1)$-forms and $F_{D-p}^i$  are $(D-p)$-forms, and $i=1,2$. Let us stress that in these expressions $D$ and $p$ are arbitrary.

One can now add masses introducing the Green--Schwarz \cite{GS} terms
\bea
- \   \star\bigg[ m_i\,B_p^i\, \wedge\, F_{D-p}^i \ \bigg]\ ,
\eea{3.2}
in a first--order form for $F_{D-p}^i$ obtained adding to the Lagrangian the terms
\bea
- \   \star\bigg[ A_p^i\, \wedge\, F_{D-p}^i \ \bigg]\  ,
\eea{3.3}
where $A_p^i=d\,C_{p-1}^i$ is $p$--form. As a result
\beq
\frac{\partial {\cal L}}{\partial F^i_{D-p}}\ = \ A^i_{p}\, +\, m_i\,B^i_{p}\ , \qquad (i=1,2)
\eeq{3.4}
and whenever a mass term $m^i$ is not vanishing one can eliminate the term involving $A_p^i$ by a gauge transformation, ending up with
\bea
{\cal L} \ &=&\ \mu^2\left[1\ -\ \sqrt{1\,+\,\frac{(H_{p+1}^1)^2+(H_{p+1}^2)^2}
{\mu^2}\,+\,\frac{(H_{p+1}^1)^2\,(H_{p+1}^2)^2-\left(H_{p+1}^1\cdot H_{p+1}^2\right)^2}
{\mu^4}}\;\right]\\ \nonumber
&+&\ \nu^2\left[1\ -\ \sqrt{1\,+\,\frac{m_1^2(B_{p}^1)^2+m_2^2(B_{p}^2)^2}
{\nu^2}\,+\,m_1^2\,m_2^2\frac{(B_{p}^1)^2\,(B_{p}^2)^2-\left(B_{p}^1\cdot B_{p}^2\right)^2}
{\nu^4}}\;\right]\ .
\eea{3.5}

In the special case $D=2 (p+1)$, one can also start from a $\mu$--dependent Lagrangian with a geometric term, and the same steps then lead to
\bea
{\cal L}\ &=&\ \mu^2\left[1\ -\ \sqrt{1+\frac{(H_{p+1}^1)^2+(H_{p+1}^2)^2}
{\mu^2}\,-\, \frac{\left(\star\left[ H^1_{p+1} \wedge H^2_{p+1}  \right]\right)^2}
{\mu^4}}\;\right]\\\nonumber
&+&\ \nu^2\left[1\ -\ \sqrt{1\,+\,\frac{m_1^2(B_{p}^1)^2+m_2^2(B_{p}^2)^2}
{\nu^2}\,+\,m_1^2\,m_2^2\frac{(B_{p}^1)^2\,(B_{p}^2)^2-\left(B_{p}^\cdot B_{p}^2\right)^2}
{\nu^4}}\;\right]\,.
\eea{3.7}
On the other hand, when $D=2\, p$  one can start from $\nu$--dependent Lagrangian with a geometric term,
obtaining
\bea
{\cal L}\ &=&\ \mu^2\left[1\ -\ \sqrt{1\,+\,\frac{(H_{p+1}^1)^2+(H_{p+1}^2)^2}
{\mu^2}\,+\,\frac{(H_{p+1}^1)^2\,(H_{p+1}^2)^2-\left(H_{p+1}^1\cdot H_{p+1}^2\right)^2}
{\mu^4}}\;\right]\nonumber\\
&+&\ \nu^2\left[1\ -\ \sqrt{1\,+\,\frac{m_1^2(B_{p}^1)^2+m_2^2(B_{p}^2)^2}
{\nu^2}\,-\, m_1^2\,m_2^2\frac{\left(\star\left[ B^1_{p} \wedge B^2_{p}  \right]\right)^2}
{\mu^4}}\;\right]\, .
\eea{3.9}

As a last example, let us consider the Lagrangian
\beq
{\cal L}\ =\ \mu^2 \,\left[\, 1\ -\ \sqrt{1\, +\, \frac{X}{\mu^2}\, -\, \frac{Y^2}{\mu^4}}\, \right]\,+\,\nu^2 \,\left[\, 1\ -\ \sqrt{1\, +\, \frac{X_0}{\mu^2}\, -\, \frac{Y_0^2}{\mu^4}}\, \right]\ ,
\eeq{3.10}
where $X$, $Y$ are defined in eq.~\eqref{2.2} and
\bea
X_0&=&\,F_{p}^2 \ + \ G_{D-p}^2 \ , \nonumber\\
Y_0&=& \star\bigg[ \ F_{p}\, \wedge\, G_{D-p} \ \bigg]\ .
\eea{3.11}
In this case one can introduce a single mass, introducing the Green--Schwarz \cite{GS} term
\bea
- \   \star\bigg[ m\,B_p\, \wedge \, G_{D-p} \ \bigg]\ ,
\eea{3.12}
in a first--order form for $G_{D-p}$ that can be reached adding to the Lagrangian the term
\bea
- \   \star\bigg[ A_p\, \wedge\, G_{D-p} \ \bigg]\  ,
\eea{3.13}
where $A_p=d\,C_{p-1}$ is $p$--form. As a result
\beq
\frac{\partial {\cal L}}{\partial G_{D-p}}\ = \ A_{p}\, +\, m\,B_{p} \ ,
\eeq{3.14}
and whenever the mass term $m$ is not vanishing one can eliminate the term involving $A_p$ by a gauge transformation, ending up with
\bea
{\cal L}&=& \mu^2 \,\left[\, 1\ -\ \sqrt{1\, +\, \frac{X}{\mu^2}\, -\, \frac{Y^2}{\mu^4}}\, \right]\nonumber\\
&+& \nu^2 \,\left[\, 1\ -\ \sqrt{1\,+\,\frac{(F_{p})^2+m^2\,(B_{p})^2}
{\nu^2}\,+\,m^2\,\frac{(F_{p})^2\,(B_{p})^2-\left(F_{p}\cdot B_{p}\right)^2}
{\nu^4}}\, \right]\ .
\eea{3.15}
%
%%%%%%%%%%%%%%%%%%%%%%%%%%
\section{Concluding remarks}\label{sec:conclusions}

This paper was motivated by intriguing analogies among different of non--linear Lagrangians for $N=2 \to N=1$ partial supersymmetry breaking in four dimensions introduced in \cite{BG,BG2,RT}. These realizations differ in the nature of the supermultiplet that hosts the goldstino mode of the broken supersymmetry, and yet the two cases of the vector multiplet and of the tensor (linear) multiplet rest on superfields strengths of opposite chiralities subject to similar non--linear constraints. This fact has direct implications for the mathematical structure of the corresponding non--linear Lagrangians, especially when they are formulated in terms of auxiliary fields, as emphasized in the Introduction. Moreover, a standard duality between linear and chiral multiplets converts the non--linear Lagrangian for the tensor multiplet into a Nambu--Goto Lagrangian for a chiral multiplet. This state of affairs affords direct generalizations in $D$ dimensions for pairs of form field strengths of complementary degrees $p+1$ and $D-p-1$. These systems enjoy a double duality when these forms are interchanged, while they acquire a manifest $U(1)$ symmetry after a single duality turns them into systems for pairs of forms of the same degree. Additional duality properties are present for $D=2(p+1)$, when the two original forms have the same degree. In these cases the field equations of the original two--form system in eq.~\eqref{B1} acquire a continuous $U(1)$ electric--magnetic duality for $p$ odd, or a $U(1) \times U(1)$ duality, where the first factor is electric and the second is electric--magnetic, for $p$ even. For $p=1$ and $D=4$, one is thus led to a two--field Born--Infeld action that admits a $U(1)$ electric--magnetic duality that interchanges the two fields. For $p$ odd, we also presented in eq.~\eqref{B13} a different two--field non--linear Lagrangian, a complexification of the Born--Infeld theory that admits an $SU(2)$ duality. The final section was devoted to the massive deformations induced in these models by four--dimensional Green--Schwarz \cite{GS} terms, along the lines of \cite{FS}.
These patterns of continuous dualities and the ensuing non--linear constraints provide explicit realizations in non--linear systems of the general framework for duality rotations proposed by Gaillard and Zumino in \cite{GZ}.

%%%%%%%%%%%%%%%%%%%%%%%%%%%%%
\subsection*{Acknowledgements} It is a pleasure to thank P.~Aschieri for stimulating discussions. A.~S. is on sabbatical leave, supported in part by Scuola Normale Superiore and by INFN (I.S. Stefi). The authors would like to thank the CERN Ph--Th Unit for the kind hospitality.
%%%%%%%%%%%%%%%%%%%%%%%%%%%%%%%%


\begin{thebibliography}{666}

\bibitem{HLP}
J.~Hughes and J.~Polchinski,
  %``Partially Broken Global Supersymmetry and the Superstring,''
  Nucl.\ Phys.\ B {\bf 278} (1986) 147;
  %%CITATION = NUPHA,B278,147;%%
J.~Hughes, J.~Liu and J.~Polchinski,
  %``Supermembranes,''
  Phys.\ Lett.\ B {\bf 180} (1986) 370.
  %%CITATION = PHLTA,B180,370;%%

\bibitem{BG}
J.~Bagger and A.~Galperin,
  %``A New Goldstone multiplet for partially broken supersymmetry,''
  Phys.\ Rev.\ D {\bf 55} (1997) 1091
  [hep-th/9608177].
  %%CITATION = HEP-TH/9608177;%%

\bibitem{BG2}
J.~Bagger and A.~Galperin,
  %``The Tensor Goldstone multiplet for partially broken supersymmetry,''
  Phys.\ Lett.\ B {\bf 412} (1997) 296
  [hep-th/9707061].
  %%CITATION = HEP-TH/9707061;%%

\bibitem{RT}
M.~Rocek and A.~A.~Tseytlin,
  %``Partial breaking of global D = 4 supersymmetry, constrained superfields, and three-brane actions,''
  Phys.\ Rev.\ D {\bf 59} (1999) 106001
  [hep-th/9811232].
  %%CITATION = HEP-TH/9811232;%%

\bibitem{deser}
S.~Deser and R.~Puzalowski,
  %``Supersymmetric Nonpolynomial Vector Multiplets and Causal Propagation,''
  J.\ Phys.\ A {\bf 13} (1980) 2501.
  %%CITATION = JPAGA,A13,2501;%%

\bibitem{CF}
S.~Cecotti and S.~Ferrara,
  %``Supersymmetric Born-infeld Lagrangians,''
  Phys.\ Lett.\ B {\bf 187} (1987) 335.
  %%CITATION = PHLTA,B187,335;%%

\bibitem{BI}
 M.~Born and L.~Infeld,
  %``Foundations of the new field theory,''
  Proc.\ Roy.\ Soc.\ Lond.\ A {\bf 144} (1934) 425.
  %%CITATION = PRSLA,A144,425;%%

\bibitem{nilpotent}
M.~Rocek,
  %``Linearizing the Volkov-Akulov Model,''
  Phys.\ Rev.\ Lett.\  {\bf 41} (1978) 451;
  %%CITATION = PRLTA,41,451;%%
U.~Lindstrom and M.~Rocek,
  %``Constrained Local Superfields,''
  Phys.\ Rev.\ D {\bf 19}, 2300 (1979);
  %%CITATION = PHRVA,D19,2300;%%
R.~Casalbuoni, S.~De Curtis, D.~Dominici, F.~Feruglio and R.~Gatto,
  %``Nonlinear Realization of Supersymmetry Algebra From Supersymmetric Constraint,''
  Phys.\ Lett.\ B {\bf 220} (1989) 569;
  %%CITATION = PHLTA,B220,569;%%
Z.~Komargodski and N.~Seiberg,
  %``From Linear SUSY to Constrained Superfields,''
  JHEP {\bf 0909} (2009) 066
  [arXiv:0907.2441 [hep-th]].
  %%CITATION = ARXIV:0907.2441;%%

\bibitem{ADFS}
I.~Antoniadis, E.~Dudas, S.~Ferrara and A.~Sagnotti,
  %``The Volkov-Akulov-Starobinsky supergravity,''
  Phys.\ Lett.\ B {\bf 733} (2014) 32
  [arXiv:1403.3269 [hep-th]].
  %%CITATION = ARXIV:1403.3269;%%

\bibitem{FKL}
  S.~Ferrara, R.~Kallosh and A.~Linde,
  %``Cosmology with Nilpotent Superfields,''
  JHEP {\bf 1410} (2014) 143
  [arXiv:1408.4096 [hep-th]];
  %%CITATION = ARXIV:1408.4096;%%

\bibitem{KL}
R.~Kallosh and A.~Linde,
  %``Inflation and Uplifting with Nilpotent Superfields,''
  JCAP {\bf 1501} (2015) 01,  025
  [arXiv:1408.5950 [hep-th]].
  %%CITATION = ARXIV:1408.5950;%%

\bibitem{DZ}
  G.~Dall'Agata and F.~Zwirner,
  %``On sgoldstino-less supergravity models of inflation,''
  JHEP {\bf 1412} (2014) 172
  [arXiv:1411.2605 [hep-th]].
  %%CITATION = ARXIV:1411.2605;%%

\bibitem{bsb}
S.~Sugimoto,
%``Anomaly cancellations in type I D9-D9-bar system and the USp(32)  string
%theory,''
Prog.\ Theor.\ Phys.\  {\bf 102} (1999) 685 [arXiv:hep-th/9905159];
%%CITATION = HEP-TH 9905159;%%
I.~Antoniadis, E.~Dudas and A.~Sagnotti,
%``Brane supersymmetry breaking,''
Phys.\ Lett.\ {\bf B 464} (1999) 38 [arXiv:hep-th/9908023];
%%CITATION = HEP-TH 9908023;%%
C.~Angelantonj,
%``Comments on open-string orbifolds with a non-vanishing B(ab),''
Nucl.\ Phys.\ {\bf B 566} (2000) 126 [arXiv:hep-th/9908064];
%%CITATION = HEP-TH 9908064;%%
G.~Aldazabal and A.~M.~Uranga,
%``Tachyon-free non-supersymmetric type IIB orientifolds via  brane-antibrane
%systems,''
JHEP {\bf 9910} (1999) 024 [arXiv:hep-th/9908072];
%%CITATION = HEP-TH 9908072;%%
C.~Angelantonj, I.~Antoniadis, G.~D'Appollonio, E.~Dudas and
A.~Sagnotti,
%``Type I vacua with brane supersymmetry breaking,''
Nucl.\ Phys.\ {\bf B 572} (2000) 36 [arXiv:hep-th/9911081].
%%CITATION = HEP-TH 9911081;%%

\bibitem{dm}
E.~Dudas and J.~Mourad,
 %``Consistent gravitino couplings in non-supersymmetric strings,''
 Phys.\ Lett.\  {\bf B 514} (2001) 173
 [arXiv:hep-th/0012071];
 %%CITATION = PHLTA,B514,173;%%
G.~Pradisi and F.~Riccioni,
 %``Geometric couplings and brane supersymmetry breaking,''
 Nucl.\ Phys.\  {\bf B 615}, 33 (2001)
 [arXiv:hep-th/0107090].
 %%CITATION = NUPHA,B615,33;%%

\bibitem{KKLT}
S.~Kachru, R.~Kallosh, A.~D.~Linde and S.~P.~Trivedi,
  %``De Sitter vacua in string theory,''
  Phys.\ Rev.\ D {\bf 68} (2003) 046005
  [hep-th/0301240];
  %%CITATION = HEP-TH/0301240;%%
S.~Kachru, R.~Kallosh, A.~D.~Linde, J.~M.~Maldacena, L.~P.~McAllister and S.~P.~Trivedi,
  %``Towards inflation in string theory,''
  JCAP {\bf 0310} (2003) 013
  [hep-th/0308055].
  %%CITATION = HEP-TH/0308055;%%

\bibitem{KW}
R.~Kallosh and T.~Wrase,
  %``Emergence of Spontaneously Broken Supersymmetry on an Anti-D3-Brane in KKLT dS Vacua,''
  JHEP {\bf 1412} (2014) 117
  [arXiv:1411.1121 [hep-th]].
  %%CITATION = ARXIV:1411.1121;%%

\bibitem{KvP}
R.~Kallosh, A.~Linde, B.~Vercnocke and T.~Wrase,
  %``Analytic Classes of Metastable de Sitter Vacua,''
  JHEP {\bf 1410} (2014) 11
  [arXiv:1406.4866 [hep-th]];
  %%CITATION = ARXIV:1406.4866;%%
R.~Kallosh, A.~Linde and M.~Scalisi,
  %``Inflation, de Sitter Landscape and Super-Higgs effect,''
  arXiv:1411.5671 [hep-th];
  %%CITATION = ARXIV:1411.5671;%%
  E.~A.~Bergshoeff, K.~Dasgupta, R.~Kallosh, A.~Van Proeyen and T.~Wrase,
  %``$\overline{\rm D3}$ and dS,''
  arXiv:1502.07627 [hep-th].
  %%CITATION = ARXIV:1502.07627;%%

\bibitem{kst}
S.~M.~Kuzenko and S.~Theisen,
  %``Nonlinear selfduality and supersymmetry,''
  Fortsch.\ Phys.\  {\bf 49} (2001) 273
  [hep-th/0007231].
  %%CITATION = HEP-TH/0007231;%%

\bibitem{ADT}
L.~Andrianopoli, R.~D'Auria and M.~Trigiante,
  %``On the dualization of Born-Infeld theories,''
  arXiv:1412.6786 [hep-th].
  %%CITATION = ARXIV:1412.6786;%%

\bibitem{FPS}
S.~Ferrara, M.~Porrati and A.~Sagnotti,
  %``N = 2 Born-Infeld attractors,''
  JHEP {\bf 1412} (2014) 065
  [arXiv:1411.4954 [hep-th]].
  %%CITATION = ARXIV:1411.4954;%%

\bibitem{FPSSY}
S.~Ferrara, M.~Porrati, A.~Sagnotti, R.~Stora and A.~Yeranyan,
  %``Generalized Born--Infeld Actions and Projective Cubic Curves,''
  arXiv:1412.3337 [hep-th],
  %%CITATION = ARXIV:1412.3337;%%
  to appear in Fortschritte der Physik.

\bibitem{GZ}
 M.~K.~Gaillard and B.~Zumino,
  %``Duality Rotations for Interacting Fields,''
  Nucl.\ Phys.\ B {\bf 193} (1981) 221; M.~K.~Gaillard and B.~Zumino,
  %``Selfduality in nonlinear electromagnetism,''
  Lect.\ Notes Phys.\  {\bf 509} (1998) 121
  [hep-th/9705226]. For a recent review see:
  P.~Aschieri, S.~Ferrara and B.~Zumino,
  %``Duality Rotations in Nonlinear Electrodynamics and in Extended Supergravity,''
  Riv.\ Nuovo Cim.\  {\bf 31} (2008) 625
  [arXiv:0807.4039 [hep-th]].


\bibitem{ONN}
S.~Cecotti, S.~Ferrara and L.~Girardello,
  %``Hidden Noncompact Symmetries in String Theory,''
  Nucl.\ Phys.\ B {\bf 308} (1988) 436;
  %%CITATION = NUPHA,B308,436;%%
  L.~Andrianopoli, R.~D'Auria and S.~Ferrara,
  %``U duality and central charges in various dimensions revisited,''
  Int.\ J.\ Mod.\ Phys.\ A {\bf 13} (1998) 431
  [hep-th/9612105].
  %%CITATION = HEP-TH/9612105;%%
  E.~Cremmer, B.~Julia, H.~Lu and C.~N.~Pope,
  %``Dualization of dualities. 1.,''
  Nucl.\ Phys.\ B {\bf 523} (1998) 73
  [hep-th/9710119],
  %%CITATION = HEP-TH/9710119;%%
  %``Dualization of dualities. 2. Twisted self-duality of doubled fields, and superdualities,''
  Nucl.\ Phys.\ B {\bf 535} (1998) 242
  [hep-th/9806106].
  %%CITATION = HEP-TH/9806106;%%

\bibitem{AGW}
L.~Alvarez-Gaume and E.~Witten,
  %``Gravitational Anomalies,''
  Nucl.\ Phys.\ B {\bf 234} (1984) 269.
  %%CITATION = NUPHA,B234,269;%%

\bibitem{orientifolds}
A.~Sagnotti, in Cargese '87, ``Non-Perturbative Quantum Field
Theory'', eds. G. Mack et al (Pergamon Press, 1988), p. 521,
%``Open Strings And Their Symmetry Groups,''
arXiv:hep-th/0208020;
%%CITATION = HEP-TH 0208020;%%
G.~Pradisi and A.~Sagnotti,
%``Open String Orbifolds,''
Phys.\ Lett.\ {\bf B 216} (1989) 59;
%%CITATION = PHLTA,B216,59;%%
P.~Horava,
%``Strings On World Sheet Orbifolds,''
Nucl.\ Phys.\ {\bf B 327} (1989) 461,
%%CITATION = NUPHA,B327,461;%%
%``Background Duality Of Open String Models,''
Phys.\ Lett.\ {\bf B 231} (1989) 251;
%%CITATION = PHLTA,B231,251;%%
M.~Bianchi and A.~Sagnotti,
%``On The Systematics Of Open String Theories,''
Phys.\ Lett.\ {\bf B 247} (1990) 517,
%%CITATION = PHLTA,B247,517;%%
%``Twist Symmetry And Open String Wilson Lines,''
Nucl.\ Phys.\ {\bf B 361} (1991) 519;
%%CITATION = NUPHA,B361,519;%%
M.~Bianchi, G.~Pradisi and A.~Sagnotti,
%``Toroidal compactification and symmetry breaking in open string theories,''
Nucl.\ Phys.\ {\bf B 376} (1992) 365;
%%CITATION = NUPHA,B376,365;%%
A.~Sagnotti,
 %``A Note on the Green-Schwarz mechanism in open string theories,''
 Phys.\ Lett.\  {\bf B 294} (1992) 196
 [arXiv:hep-th/9210127]~.
 %%CITATION = PHLTA,B294,196;%%
For reviews see: E.~Dudas,
%``Theory and phenomenology of type I strings and M-theory,''
Class.\ Quant.\ Grav.\  {\bf 17} (2000) R41 [arXiv:hep-ph/0006190];
%%CITATION = HEP-PH 0006190;%%
C.~Angelantonj and A.~Sagnotti,
%``Open strings,''
Phys.\ Rept.\  {\bf 371} (2002) 1 [Erratum-ibid.\  {\bf 376} (2003)
339] [arXiv:hep-th/0204089].
%%CITATION = HEP-TH 0204089;%%

\bibitem{orientifolds2}
S.~Ferrara, R.~Minasian and A.~Sagnotti,
  %``Low-energy analysis of M and F theories on Calabi-Yau threefolds,''
  Nucl.\ Phys.\ B {\bf 474} (1996) 323
  [hep-th/9604097];
  %%CITATION = HEP-TH/9604097;%%
S.~Ferrara, F.~Riccioni and A.~Sagnotti,
  %``Tensor and vector multiplets in six-dimensional supergravity,''
  Nucl.\ Phys.\ B {\bf 519} (1998) 115
  [hep-th/9711059].
  %%CITATION = HEP-TH/9711059;%%

\bibitem{ABMZ}
P.~Aschieri, D.~Brace, B.~Morariu and B.~Zumino,
  %``Nonlinear selfduality in even dimensions,''
  Nucl.\ Phys.\ B {\bf 574} (2000) 551
  [hep-th/9909021].
  %%CITATION = HEP-TH/9909021;%%

\bibitem{GS}
M.~B.~Green and J.~H.~Schwarz,
  %``Anomaly Cancellation in Supersymmetric D=10 Gauge Theory and Superstring Theory,''
  Phys.\ Lett.\ B {\bf 149} (1984) 117.
  %%CITATION = PHLTA,B149,117;%%

\bibitem{bkky}
S.~Bellucci, N.~Kozyrev, S.~Krivonos and A.~Yeranyan,
  %``Supermembrane in $D = 5$: component action,''
  JHEP {\bf 1405} (2014) 142
  [arXiv:1312.0231 [hep-th]].
  %%CITATION = ARXIV:1312.0231;%%

\bibitem{FS}
S.~Ferrara and A.~Sagnotti,
  %``Massive Born--Infeld and Other Dual Pairs,''
  arXiv:1502.01650 [hep-th],
  %%CITATION = ARXIV:1502.01650;%%
to appear in JHEP.

\end{thebibliography}
\end{document}